\documentclass[%
 reprint,
 amsmath,amssymb,
 aps,pre,
floatfix,
]{revtex4-2}

\usepackage{graphicx}
\usepackage{dcolumn}
\usepackage{bm}

\usepackage{hyperref}
\hypersetup{colorlinks,allcolors=black}

\usepackage{bm}

\usepackage{xcolor}
\usepackage{amsbsy}
\usepackage{subfigure}
\usepackage{tikz}
\usepackage{footnote}
\begin{document}

\title{Two fitness inference schemes compared using allele frequencies from 1,068,391 sequences sampled in the UK during the COVID-19 pandemic}



\author{Hong-Li Zeng}
\author{Cheng-Long Yang}
\author{Bo Jing}
 \affiliation{School of Science, Nanjing University of Posts and Telecommunications, Key Laboratory of Radio and Micro-Nano Electronics of Jiangsu Province, Nanjing, 210023, CHINA}
\email{hlzeng@njupt.edu.cn}
\author{John Barton}%
\affiliation{%
 Department of Computational \& Systems Biology, University of Pittsburgh School of Medicine, USA}%
\author{Erik Aurell}
\affiliation{
 Department of Computational Science and Technology, AlbaNova University Center, SE-106 91 Stockholm, SWEDEN}%
\email{eaurell@kth.se}%

\date{\today}

\begin{abstract}
Throughout the course of the SARS-CoV-2 pandemic, genetic variation has contributed to the spread and persistence of the virus. 
For example, various mutations have allowed SARS-CoV-2 to escape antibody neutralization or to bind more strongly to the receptors that it uses to enter human cells.
Here, we compared two methods that estimate the fitness effects of viral mutations using the abundant sequence data gathered over the course of the pandemic. Both approaches are grounded in population genetics theory but with different assumptions. One approach, tQLE, features an epistatic fitness landscape and assumes that alleles are nearly in linkage equilibrium. Another approach, MPL, assumes a simple, additive fitness landscape, but allows for any level of correlation between alleles. We characterized differences in the distributions of fitness values inferred by each approach and in the ranks of fitness values that they assign to sequences across time. 
We find that in a large fraction of weeks
the two methods are in good agreement as to
their top-ranked sequences, 
\textit{i.e.} as to which
sequences observed that week are most fit.
We also find that agreement between ranking of sequences varies with genetic unimodality in the population in a given week.

\begin{description}
\item[keywords]
SARS-CoV-2 $|$ allele frequency time series $|$ fitness inference $|$
transient Quasi-Linkage\\ Equilibrium (tQLE) $|$ 
marginal path likelihood (MPL)
\end{description}
\end{abstract}
\maketitle

\section{Introduction}
The COVID-19 pandemic 
had the largest impact on world-wide human health by an infectious 
disease agent since the Spanish flu more than a century ago \cite{WHO}.
After more than three years of at times high infection rates in practically all countries in the world,
the disease has reached an endemic state, and the virus will likely remain in circulation in
the foreseeable future.
The spread of SARS-CoV-2 was accompanied by the emergence of many variants, 
some of which successfully replaced earlier variants. These variants 
differed in their virulence, infectiousness, and resistance to vaccines.
They also differed in their exact genotypes, as determined by many
high-quality whole-genome sequences deposited in repositories such as 
GISAID \cite{GISAID}.

The unprecedented amount of genomic time series data collected for SARS-CoV-2 allows for
analysis that was previously impossible. In particular, this data enables the development and comparison of prediction and/or inference methods that may be useful in a future pandemic, an event that is likely unavoidable even if challenging to predict.
Genomic time series analysis also allows for feature discovery, which can help shed light on the biology of the virus in its newly conquered environment, i.e.~the human population.

We will here compare and contrast two recently developed approaches for fitness inference from genetic time series data. One approach is based on the quasi-linkage equilibrium (QLE) theory of
Kimura \cite{Kimura1965} and Neher and Shraiman \cite{NeherShraiman2009, NeherShraiman2011}, which we will here use in a dynamic (non-stationary) version
which we call
tQLE. The basic idea of tQLE is to infer parameters of models in exponential families describing the distribution of genotypes in a population from sequence data with time stamps. Due to the high sampling world-wide during the pandemic, sequence data can be sampled precisely in time, down to periods of even a single week. In this approach, the SARS-CoV-2 genotype distribution is first described by Potts parameters $h_i(t)$ and $J_{ij}(t)$ where $t$ is the sample time (metadata available in GISAID), and where the data can optionally also be stratified by the region of origin for each sample. QLE theory 
relates epistatic fitness [parameters $f_{ij}$] to Potts parameters $J_{ij}$.
tQLE additionally gives a relationship between the contribution of additive fitness from variation at genomic position $i$ [parameter $f_i$], Potts parameters
$h_i$ and $J_{ij}$, and the time derivative $\dot{h}_i$.

The second approach called marginal path likelihood (MPL) was recently developed by Barton and co-workers \cite{Sohail2021}, and applied by them on SARS-CoV-2 data up to August 2021 \cite{Lee2021}. The main idea of MPL is to estimate the probability of an observed history of allele frequencies from a Wright-Fisher model (or, in the case of SARS-CoV-2, a branching process epidemiological model \cite{Lee2021}), including recombination, and then maximize this probability over model parameters. The resulting formula involves frequencies, mutational pressure, and linkage disequilibrium (correlation) between alleles. This approach has some similarities to an inference formula from time series for neuroscience applications developed by two of us some time ago \cite{Zeng2013}. 

The main conceptual 
difference between the two methods is that
MPL is derived from the finite-$N$
noisy dynamics of single-locus frequencies
while tQLE is derived from the
infinite-$N$ noise-less dynamics of
single-locus frequencies and two-locus pair frequencies.
tQLE allows for both additive
and pairwise epistatic contributions
to fitness, if the additive
contribution dominates, so that
the instantaneous distribution is
close to linkage equilibrium.
A scenario when this
happens, and which is assumed in the version of tQLE used in this work, is 
when recombination is
a faster process than selection and
mutations
\cite{NeherShraiman2011,Zeng2020}.
Other scenarios are however also possible \cite{Dichio2021}, and could be
used as a basis for other variants of tQLE.
The version of
MPL used here is derived from single-locus 
frequencies, and it can only be used
to infer additive fitness.
On the other hand, the incorporation
of stochastic (finite-$N$) 
effects and not requiring the assumption of
quasi-linkage equilibrium are advantages
of MPL. 

Here, we compared the fitness values inferred by tQLE and MPL for weekly batches of SARS-CoV-2 sequences, collected over the first few years of the pandemic.
We found good agreement between the two methods on the relative ranks of sequences in terms of their fitness.
The similarity of the rankings is especially notable given the differences in modeling assumptions for tQLE and MPL.

\section{Materials} 
\subsection*{Data preparation}

The data utilized here was sourced from the GISAID repository, spanning from the beginning of the COVID-19 pandemic until August 12, 2023. Subsequently, a remarkable decline in the number of genomes uploaded to the database was observed. The inclusion criteria for our analysis involved selecting only high-quality and full-length genomes, as per the defined standards outlined on the GISAID website. All retained genomes possess a length not exceeding 29,903 base pairs. Each genome is labeled based on its sample collection date, a metadata parameter provided by GISAID. Given the observable bias in alternative submission times to GISAID \cite{Kalia2021lag}, sequences are systematically stratified weekly, resulting in a total of 179 datasets encompassing 5,644,661 sequences. Due to a significant geographical imbalance in the collected samples of SARS-CoV-2, the analysis reported here  
specifically focused on the UK region.

\subsection*{Data processing}
The data for the regions in 
In Fig.~\ref{fig:num-of-seqs-per-week}
satisfy the following minimal criteria:
\begin{itemize}
    \item In any period of 5 days within the time series, there are at least 20 total samples.
    \item  The number of days in the time series is greater than 20.
\end{itemize}
Applying these 
criteria, our dataset for the UK region spans 170 weeks, ranging from 2020-03-14 to 2023-06-04. The dataset for the UK comprises 1,068,391 sequences in total.

For the sequences in each week, a Multiple Sequence Alignment was constructed through the \texttt{MAFFT} software\cite{Katoh2017, Kuraku2013}. Sequences from each week are aligned separately to the reference sequence ``Wuhan-Hu-1'', with GISAID accession number EPI\_ISL\_402125 \cite{Chen2020}.
Note that this is different from the procedure in \cite{Zeng2020pnas}, where pre-aligned MSAs were used. The total number of sequences in that study was much less than that used here. 

Each MSA is a matrix $\boldsymbol{\sigma}=\{\sigma_i^n|i=1,...,L, n = 1,...,N\}$, where $N$ represents the number of genomic sequences in a week while $L$ represents the number of loci in an aligned sequence~\cite{Cocco2018, Horta2019}. 
Thus, all aligned sequences have a length of $L=29,903$, the same as that of the reference sequence, while $N$ by construction varies from week to week, see Fig.~\ref{fig:num-of-seqs-per-week}. The loci between 256 and 29,674 are referred to as \textit{coding region}, since they code for the protein-coding genes in the
SARS-CoV-2 genome.
Each entry $\sigma_i^n$ of the MSA $\boldsymbol{\sigma}$ is either one of the $4$ nucleotides (A,C,G,T), or the alignment gap `-', the minorities like ``KYF...'' are changed to the sign of `-' for the sake of simplicity of the following allele frequency analysis. 

\subsection*{Data filtering}
In Fig.~\ref{fig:freq_UK} we show the allele frequency time series
for all loci in the UK data, and in Fig.~\ref{fig:freq-coding_region_UK}
only for loci in the coding region.
These two figures show that at a majority of loci all sequences contain the same symbol, and most of the remaining variation is in the non-coding regions. In a first filtering step we retain in the analysis loci where the 
most frequent mutation 
away from wild-type is classified as ``Non-synonymous mutation'' as defined in \cite{Sohail2021}
and where the largest mutant frequency
is at least 1\%.
In a second step we retain only those
loci which meet the criteria for all weeks. For the UK data used in this study there remains 209 loci.

For the other data sets
shown in 
Fig.~\ref{fig:num-of-seqs-per-week}
(lower panel) 
there would remain respectively 
1063 loci in ``Global'',
173 loci in ``EU'', 328 loci in ``NA'' 
(North America) and 225 loci in ``Asia''.
For consistency, the average Hamming distances are however 
for all computed from the variability at the same 209 loci as in the UK data set.

\section{Methods}

\subsection*{The driving forces of evolution: selection, mutation, recombination and genetic drift}
In both approaches to be considered, the driving forces of evolution are assumed (Darwinian) selection, mutation, recombination, and genetic drift (finite population effects). Effects excluded from consideration are hence e.g. spatial barriers (island models). The genome $\mathbf{x}=\left(x_1,x_2,\ldots,x_L\right)$ where each $x_i$ is an indicator variable of the allele (nucleotide in the set $\{-,N,A,C,G,T\}$ at locus $i$ (position $i$ in the MSA).

\paragraph{Fitness} is assumed to be a function
\begin{equation}
\label{eq:def-fitness-function}
    F(\mathbf{x}) = \sum_{i=1}^L\sum_a f^{(1)}_{i,a} \mathbf{1}_{x_i,a}
    + \sum_{i,j=1}^L\sum_{a,b} f^{(2)}_{ij,ab} \mathbf{1}_{x_i,a} \mathbf{1}_{x_j,b}
\end{equation}
The coefficients $f^{(1)}_{i,a}$ are called \textit{additive fitness} and 
parameterize the selective advantage of allele $a$ at locus $i$ with respect to wild-type.
The coefficients $f^{(2)}_{ij,ab}$ are called (pair-wise) \textit{epistatic fitness} and 
parameterize the selective advantage of alleles $a$ and $b$ at loci $i$ and $j$ beyond what they contribute separately. 
In tQLE $f^{(2)}_{ij,ab}$ are adjustable parameters inferred from the data, which for self-consistency however cannot be too large. In MPL $f^{(2)}_{ij,ab}$ are absent.

The evolution of genotypes in the population over a short time $\Delta t$ due to fitness is on the level of a normalized distribution given by
\begin{equation}
\label{eq:def-fitness}
    P(\mathbf{x},t+\Delta t)|_{\mathrm{fitness}} = \frac{e^{\Delta t F(\mathbf{x})}}
    {\sum_{\mathbf{x}} e^{\Delta t F(\mathbf{x})} P(\mathbf{x},t)} P(\mathbf{x},t)
\end{equation}
In both approaches our goal is to estimate the coefficients $f^{(1)}_{i,a}$.

\paragraph{Mutations}
are random changes of single alleles. In general, they could be parametrized
acting as
\begin{eqnarray}
\label{eq:def-mutations}
    P(\mathbf{x},t+\Delta t)|_{\mathrm{mutation}} &=& 
    P(\mathbf{x},t) + \Delta t \sum_{i}\sum_{ab}  \mathbf{1}_{x_i,a}
    \nonumber \\ && \qquad \big( \mu_i^{ba} P(M_i^{ab}\mathbf{x},t)
     -  \mu_i^{ab} P(\mathbf{x},t) \big)
\end{eqnarray}
where $M_i^{ab}$ is the flip operator which changes allele $a$ at locus $i$ to $b$, and 
$\mu_i^{ab}$ rate of this process. These rates are only partially known, and only parametrise a fraction of naturally occurring mutations. As our focus is here on fitness we will follow the original theoretical literature \cite{NeherShraiman2011,Sohail2021} and take them all equal to one overall mutation rate $\mu$. 

\paragraph{Recombination} is modelled as a process whereby two genomes combine and give rise to a third. On the level of distributions that is given by 
\begin{eqnarray}
\label{eq:def-recombination}
    &&P(\mathbf{x},t+\Delta t)|_{\mathrm{recomb}} = 
    P(\mathbf{x},t)\left(1-r \Delta t\right) + 
    \nonumber \\
    && \qquad r\Delta t \sum_{\mathbf{x}_m,\mathbf{x}_f} C(\mathbf{x};\mathbf{x}_m,\mathbf{x}_f) P(\mathbf{x}_m,t) P(\mathbf{x}_f,t)
\end{eqnarray}
In above $r$ is an overall recombination rate with dimension inverse time.
The function $C(\mathbf{x};\mathbf{x}_m,\mathbf{x}_f)$ is the specific rate at which
genomes $\mathbf{x}_m$ and $\mathbf{x}_f$ combine to yield $\mathbf{x}$.
In tQLE it only enters through the derived quantity $c_{ij}$ which is the probability
that alleles at loci $i$ and $j$ are inherited from different parents \cite{NeherShraiman2011,Gao2019,Dichio2021}. In MPL recombination drives the evolution and influences the distribution of evolutionary trajectories, but does not directly enter in the inference formulae. The factor $\left(1-r \Delta t\right)$ serves to normalize the distribution. More general models of recombination in the same context are discussed in \cite{Gao2019}. 

\paragraph{Genetic drift} is the term of stochastic effects due to a finite population. All three evolution equations \eqref{eq:def-fitness},
\eqref{eq:def-mutations} and \eqref{eq:def-recombination} are valid on the ensemble level, and can be simulated by evolving several populations in parallel, and then averaging. Single-locus frequencies and two-loci pair frequencies (and other characteristics) will evolve due to both deterministic drift and random noise. In QLE the corresponding stochastic differential equations for single- and two-loci frequencies are derived and discussed 
\cite{NeherShraiman2011}. In MPL the stochastic differential equations 
for single-locus frequencies are central in deriving the path probabilities
as discussed below.

\subsection*{Quasi-linkage equilibrium (QLE)}
The phase of quasi-linkage equilibrium (QLE) was discovered by Kimura in the study of a two-locus biallelic model \cite{Kimura1965}. The extension to many loci was investigated by Neher and Shraiman\cite{NeherShraiman2009,NeherShraiman2011}. The generalization to more than two alleles per locus was given in \cite{Gao2019}. The two defining properties of QLE (formalized in \cite{Dichio2021}) are
\begin{enumerate}
    \item Multi-genome probability distributions factorize such that 
    $P_n(\mathbf{x}^{(1)},\mathbf{x}^{(2)},\ldots,\mathbf{x}^{(n)})=
    P(\mathbf{x}^{(1)})P(\mathbf{x}^{(2)})\cdots P(\mathbf{x}^{(n)})$.
    This property is especially important for $n=2$ as it allows to model
    the effects of recombination as a molecular collision in kinetic gas theory
    (Boltzmann's \textit{Stosszahlansatz}). 
    \item The single-genome probability distributions are Gibbs distribution with terms no higher than in fitness. For \eqref{eq:def-fitness-function} this means the Ising-Potts distributions of equilibrium statistical mechanics
    \begin{eqnarray}
        P(\mathbf{x})&=&\frac{1}{Z(\{h\},\{J\})}
        \exp\big(\sum_{i,a} h_i(a)\mathbf{1}_{x_i,a} 
        \nonumber \\ 
        \label{eq:Ising-Potts}
        &&\qquad +\sum_{ij,ab} J_{ij}(a,b) \mathbf{1}_{x_i,a}\mathbf{1}_{x_j,b}\big)
    \end{eqnarray}
\end{enumerate}
Relations between fitness parameters $\{f^{(1)}\}$ and $\{f^{(2)}\}$  in \eqref{eq:def-fitness-function} and stationary Ising/Potts model parameters 
$(\{h\},\{J\})$ \eqref{eq:Ising-Potts} are key quantitative results of
QLE theory. In \cite{Kimura1965} and \cite{NeherShraiman2011}
they were derived in the limit where overall recombination rate $r$ is larger than both overall mutation rate $\mu$ and variations in fitness.
Direct tests using scatter plots were first given in \cite{Zeng2020}.
Several alternative relations were derived for larger mutation rates in  
\cite{Zeng-Mauri-2021}, of which one was tested in \cite{Zeng-Mauri-2021}, see also 
\cite{Dichio2021}.

\subsection*{Transient QLE (tQLE) and fitness inference from time series data}
QLE in fact is a dynamic theory where parameters
$(\{h\},\{J\})$ in general change in time. We here introduce
the derived abbreviation tQLE to emphasize that we use the formulas for inference in this regime, and not as previously where $(\{h\})$ and $(\{J\})$ are both in steady state.

The equation for $(\{J\})$ is of the relaxation type, and in the theory of \cite{NeherShraiman2011} 
\begin{equation}
    \dot{J}_{ij}(a,b) = f^{(2)}_{ij}(a,b) - r c_{ij} J_{ij}(a,b)
\end{equation}
For large enough $r$ the Potts parameters will hence relax to a stable fixed point, \textit{i.e.} to $J^*_{ij}(a,b) = \frac{f^{(2)}_{ij}(a,b)}{r c_{ij}}$, which allows to \textit{infer} epistatic fitness parameters from Potts parameters computed from the data through the formula
\begin{equation}
    \label{eq:f-2-inference}
    f^{(2),*}_{ij}(a,b) = r c_{ij} J^*_{ij}(a,b)
\end{equation}
This relation was derived in \cite{NeherShraiman2011}, and tested (in stationary state) in \cite{Zeng2020}.
As discussed in \cite{Zeng2022}
since \eqref{eq:f-2-inference} only
relates pair-wise quantities, it can also
work when the single-nucleotide frequencies
change. This could for instance be the case 
of additive fitness changes in time, say
by a change of the fitness landscape of which one
example could be the introduction of widespread
vaccination against SARS-CoV-2 in the COVID19 pandemic.

The equation for $(\{h\})$ is on the other hand not of the relaxation type
(\cite{NeherShraiman2011}, Eq.~24)
\begin{equation}
    \dot{h}_{i}(a) = f^{(1)}_{i}(a) + r \sum_{j,b} c_{ij} 
    J_{ij}(a,b) m_j(b)
\end{equation}
where $m_j(b)=\sum_{\mathbf{x}} P(\mathbf{x}) \mathbf{1}_{x_j,b}$ is the frequency of allele $b$ at locus $j$.
Combining \eqref{eq:f-2-inference} and inferred values of $\{h\}$ at two consecutive time intervals lead to the inference formula
\begin{eqnarray}
    f^{(1),*}_{i}(a,t) &=& \frac{1}{\Delta t}\left[h_{i}(a,t+\Delta t) -
    h_{i}(a,t)\right] 
    \nonumber \\
    \label{eq:f-1-inference-QLE}
    &&
    - \sum_{j,b}  f^{(2),*}_{ij}(a,b, t)\, m_j(b,t)
\end{eqnarray}

\subsection*{Relation to data-driven fitness measures}
The observation in Fig.~\ref{fig:scatter-tQLE},
that the contribution of the time derivative in
eq.~(\ref{eq:f-1-inference-QLE}) to fitness is
relatively small,
allows to contrast in a compact manner fitness as used here to data-driven fitness
measures. 
For simplicity of notation we consider here only
biallelic genomes. Given parameters $h^*_i$ and 
$J^*_{ij}$ (Ising parameters) inferred from data,
the data-driven fitness of a sequence is 
\begin{equation}
    \label{eq:inferred-sequence-fitness-MaxEnt}
    F^{DD}\left[{\mathbf s}\right] = \hbox{Const.} + \sum_i h^{*}_{i} s_i
    + \frac{1}{2}\sum_{ij} 
    J^{*}_{ij} s_i s_j
\end{equation}
In the theory derived above, neglecting the time derivative in
eq.~(\ref{eq:f-1-inference-QLE}), we have instead
\begin{equation}
    \label{eq:inferred-sequence-fitness-QLE} 
    F^{QLE}\left[{\mathbf s}\right] = \hbox{Const.} +
    \frac{r}{2}\sum_{ij} c_{ij} J^*_{ij}(s_i-m_i)(s_j-m_j)
\end{equation}
where $s_i=\pm 1$ and $m_i=\left<s_i\right>$.
In the approximation of one overall rate of recombination ($r$)
and cross-over probability ($c_{ij}$) uniformly equal to one half these two parameters
will not change sequence fitness order, and can hence be ignored. 
To compare further use the mean-field approximation
for site magnetization in the Ising model: 
\begin{equation}
    m_i = \tanh\left(h_i+ \sum_j J_{ij}m_j\right)
\end{equation}
which implies the naive mean-field inference formula for $h_i$:
\begin{equation}
    h^*_i =  \tanh^{-1}(m_i) - \sum_j J^*_{ij}m_j
\end{equation}
Hence, sequence fitness order according to eq.~(\ref{eq:inferred-sequence-fitness-MaxEnt}) and eq.~(\ref{eq:inferred-sequence-fitness-QLE}) differ 
(using the above approximations) by $\sum_i \tanh^{-1}(m_i) s_i$. 

\subsection*{The marginal path likelihood (MPL) method}
The marginal path likelihood (MPL) method \cite{Sohail2021} is based on the 
evolution of nucleotide frequencies in Kimura's diffusion approximation
\cite{Kimura1956,Kimura1964}. The starting point is thus the joint probability
$P(\{m\}^{(1)},\{m\}^{(2)},\ldots,\{m\}^{(L)})$ where $m^{(i)}_a$ is the frequency of allele $a$ on locus $i$, normalized as $\sum_a m^{(i)}_a = 1$.
In the diffusion approximation, this probability satisfies
a Fokker-Planck equation
\begin{equation}
    \label{eq:diffusion}
    \partial_t P
    = - \sum_{i,a} \frac{\partial}{\partial m^{(i)}_a}
    \left(u^{(i)}_a P\right)
    + \sum_{ij,ab} \frac{\partial^2}{\partial m^{(i)}_a\partial m^{(j)}_b}
    \left(D^{(ij)}_{ab} P\right)
\end{equation}
where the drift vector and diffusion matrix are given by
(\cite{Sohail2021}, Eq 6 and Eq S9 and following, notation aligned with the present presentation)
\begin{eqnarray}
u^{(i)}_a &=&  m^{(i)}_a (1-m^{(i)}_a)f^{(1)}_i(a) 
+ \mu\left(1- 2 m^{(i)}_a\right) 
\nonumber \\
    \label{eq:drift-vector}
&&\quad
+ \sum_{j,b} \left( 
m^{(ij)}_{ab} - m^{(i)}_a m^{(j)}_b\right) f^{(1)}_j(b) \\
    \label{eq:diffusion-matrix}
D^{(ij)}_{ab} &=& \{ \begin{array}{lr}
m^{(i)}_a m^{(i)}_b & i=j \\
m^{(ij)}_{ab} - m^{(i)}_b m^{(j)}_b & i\neq j
\end{array}
\end{eqnarray}
\begin{equation}
    f^{(2),*}_{ij}(a,b) = r c_{ij} J^*_{ij}(a,b)
\end{equation}
The Fokker-Planck equation \eqref{eq:diffusion} corresponds to a multidimensional Langevin equation for which the probability of a path sampled at discrete times can be estimated by standard arguments. Maximizing this path probability with a Gaussian prior leads to the central inference formula in MPL 

\begin{eqnarray}
&&f^{(1),*}_{i}(a) =
\sum_{j,b} 
\left[\sum_{k=1}^K \Delta t_k D^{(ij)}_{ab}(t_k)+\gamma\mathbf{1}_{ia,jb}\right]^{-1}_{ia,jb}
\nonumber \\
        \label{eq:f-1-inference-MPL}
&&
\left[m^{(j)}_b (t_K)-m^{(j)}_b (t_0) -\mu 
\sum_{k=1}^{K-1} \Delta t_k (1-2 m^{(j)}_b(t_k)\right] 
\end{eqnarray}
In above a time interval $[t_0,t_K]$ has been divided up in $K$ sampling
intervals and the allele frequencies ($m^{(i)}_a$) and drift and diffusion
terms (from \eqref{eq:drift-vector} and \eqref{eq:diffusion-matrix})
estimated for each. The sampling interval times are defined as
$\Delta t_k = t_{k+1}-t_k$. $\gamma$ is the width of the Gaussian prior,
and acts as a regularizer.

In \eqref{eq:f-1-inference-QLE} the inferred additive fitness depends on time
and is linear in the time derivative of one inferred Potts parameters 
$h_{i}(a)$. This parameter is in itself a (complicated) function of the 
single-nucleotide and pair-wise frequencies at that time. 
In \eqref{eq:f-1-inference-MPL} the inferred additive fitness
also depends on time, more specifically on a time interval, and is linear
in $m^{(i)}_a (t_K)-m^{(i)}_a (t_0)$, the change in all the single-nucleotide
frequencies over that interval. Pair-wise frequencies also enter in \eqref{eq:f-1-inference-MPL}, through the dependence of $D^{(ij)}_{ab}$
as in \eqref{eq:diffusion-matrix}.

\subsection*{Loss of QLE}
The QLE state is lost when
the distribution no longer fulfills the two listed criteria.
A well-studied loss channel at 
very low mutation rate and sufficiently
low recombination rate is through the emergency of clones
\cite{NeherShraiman2011,Neher2013}
These are groups of identical genomes of high-fitness related by common descent. Instead of as from one exponential model (as in QLE), the distribution of genotypes is instead a mixture of clones, with one separate distribution for each clone. There may also be only a single clone, in which case all (are most) of the genotypes are the same. It was established in 
\cite{Neher2013} that the transition occures at a recombination rate 
\begin{equation}
\label{eq:Neher2013}
    r* \approx \sigma_f 
    \sqrt{\log N}
\end{equation}
where $\sigma_f$ is the variation in fitness and $N$ is the size of the population. In an infinite population $r*$ is infinite so that the QLE phase would be absent. 
This is not a serious problem since $\sqrt{\log N}$ increases very slowly with $N$. On the other hand it points to that QLE can only exist as transient state in a finite population with strictly no mutations. The reason is that in this setting sooner or later the most fit genome takes over as a single dominating clone, see \cite{Gao2019} and \cite{Zeng2020} for a discussion. 

A second loss channel observed at higher mutation rate 
leads to a phase of ``noisy clones'' coexisting with a QLE-like state.
In \cite{Dichio2021} this new phase was named  
Non-Random Coexistence (NRC). 
The transition from QLE to NRC goes through an intermittent phase where the state of the population jumps between QLE and NRC. The dependence of the jump rates on population size $N$ was investigated in \cite{Dichio2021},
and leads to a qualitatively similar behaviour as 
\eqref{eq:Neher2013},
\textit{i.e.} that for a sufficiently large population only the NRC phase is stable.

\subsection*{Recombination in coronaviruses and in SARS-CoV-2}
SARS-CoV-2 is in classifications
such as Pango~\cite{Pango}, 
assumed to evolve by descent,
and the growth and subsequent
decay of SARS-CoV-2 variants \cite{VoC-6,Indian_variant,Tegally2021,B.1.1.7} is well-known.
This can be taken to be the standard view of SARS-CoV-2 evolution, analogous to the evolution of other viruses such as influenza.

On the other hand, coronaviruses in general exhibit recombination~\cite{LaiCavanagh1997,Graham2010,Hartenian2020,Li2020}, 
a process which has also been observed to occur in SARS-CoV-2~\cite{Choi2020,Baang2021,Gribble2021,Hensley2021,Kemp2021}.

The effectiveness and importance of recombination in the general SARS-CoV-2 viral population
has been questioned~\cite{VanInsberghe2021}. 
A complicating factor is that many viral variants seem themselves to evolve,
to split into sub-variants and perhaps to recombine \cite{Rono2021,Duerr2021}.

\subsection*{Rank order comparisons}
Given two lists of sequences ordered as to fitness, we
compare the rankings by the Spearman rank correlation coefficient. This is
computed with the Matlab function `corr()' with type argument
`Spearman'.

\section{Results}
\begin{figure}[!ht]
\centering
\includegraphics[width=0.48\textwidth]{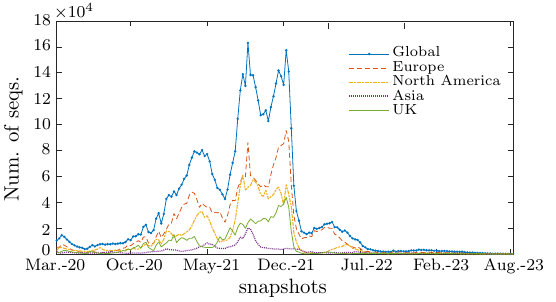}
\includegraphics[width=0.47\textwidth]{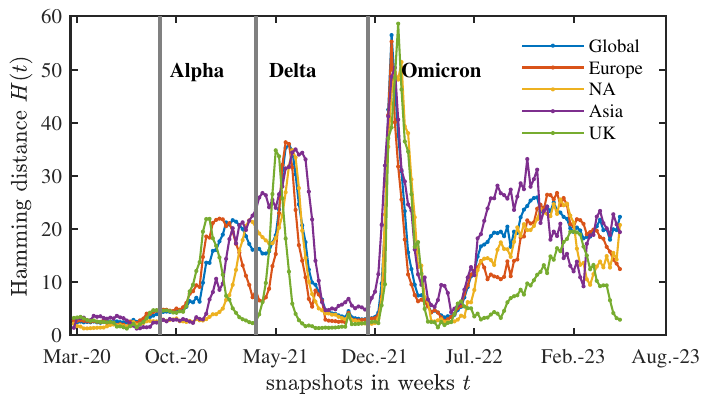}
\caption{Weekly dynamics of the number of sequences downloaded from the GISAID portal, portraying global trends (solid blue line), European contributions (dashed orange line), North American submissions (dashed yellow line), Asian datasets (purple), and those specifically from the UK excluding Ireland (solid green line). Europe and North America consistently emerge as the primary contributors to GISAID, with Europe leading in sequence submissions throughout most weeks. In 2022, Europe maintained its prominence by contributing the highest number of sequences. Lower panel: sequence diversity per week quantified by average Hamming distance between sequences collected 
and retained in the analysis that week. Vertical lines mark the first observations of Variants of Concern (VoCs) Alpha, Delta and Omicron, each of which is followed by a peak in diversity approximately when the variant rose to dominance.}
\label{fig:num-of-seqs-per-week}
\end{figure}

Globally, 5,644,661 sequences were obtained from the GISAID database. Upon weekly stratification of the data, we obtain 179 weeks in total. Notably, the sample collections of SARS-CoV-2 reveal a pronounced geographical imbalance, as depicted in Fig.~\ref{fig:num-of-seqs-per-week}. Consequently, our analysis focuses exclusively on genomic data originating from three regions in the UK (England, Wales, and Scotland). The number of sequences from the UK is 1,068,391 in total.

\subsection*{Amount and diversity of collected sequences over time} 
The weekly stratified datasets are geographically segmented for a detailed analysis. As illustrated in Fig.~\ref{fig:num-of-seqs-per-week} (upper panel), the number of sequences undergoes dynamic changes across weeks for different regions, including Europe (dashed orange lines), North America (dashed yellow lines), Asia (dashed purple line), three distinct regions of the UK (green line), and the global dataset (blue line). To account for the impact of geographic separation, data from North Ireland is intentionally excluded. Moreover, Fig.~\ref{fig:num-of-seqs-per-week} (upper panel) reveals a significant downturn in the number of collected samples in 2022, with Europe emerging as the primary source of sequences.
The diversity of collected sequences is in
Fig.~\ref{fig:num-of-seqs-per-week} (lower panel) quantified by average Hamming distance. 
One observes increased diversity after the emergence of Alpha, Delta and Omicron (marked in figure).

\begin{figure}[!ht]
\centering
\includegraphics[width=0.48\textwidth]{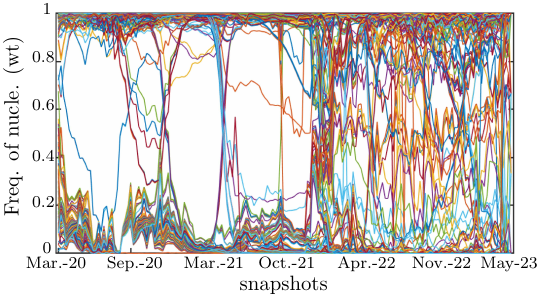}
\caption{Allele frequencies of all 29,903 sites of the nucleotides in the ``Wuhan-Hu-1'' sequence for the UK datasets over weeks. The frequencies located at the bottom mainly originate from the non-coding region (3'-UTR and 5'-UTR) of SARS-CoV-2. There are more oscillations in 2022 than those in 2020 and 2021.}
\label{fig:freq_UK}
\end{figure}

\begin{figure}[!ht]
\centering
\includegraphics[width=0.48\textwidth]{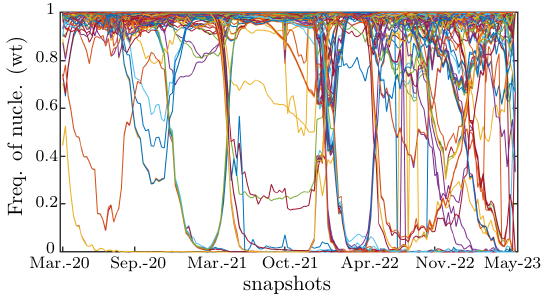}
\caption{Allele frequencies for all sites in the coding region (sites 266 to 29,674) for the UK datasets of the wild-type nucleotides from the ``Wuhan-Hu-1'' sequence. The fluctuations as shown at the bottom of Fig.~\ref{fig:freq_UK} disappear here. The oscillations depicted here are more visible than those in Fig.~\ref{fig:freq_UK}.}
\label{fig:freq-coding_region_UK}
\end{figure}
\subsection*{Allele Frequencies}
We computed allele frequencies over time from SARS-CoV-2 multiple sequence alignments from the UK. Subsequently, the allele frequencies for the nucleotides in the ``Wuhan-Hu-1'' reference sequence are selected. Fig.~\ref{fig:freq_UK} illustrates the allele frequencies over all loci ($L=29,903$ base pairs in total), while Fig.~\ref{fig:freq-coding_region_UK} specifically shows the allele frequencies within the coding region, 
spanning from the 256th to the 29,674th sites in the sequence.
There are sustained fluctuations at the bottom of Fig.~\ref{fig:freq_UK}, 
which hence
mainly originate from the non-coding region (3$^\prime$-UTR and 5$^\prime$-UTR parts) of SARS-CoV-2. Both plots display variations at specific loci within the coding region, 
of which many of them can be related to listed mutations in known Variants of Concern,
compare monthly
data and a relation to Omicron reported in 
\cite{Zeng2021b} and \cite{Zeng2022}
(Fig.~5).
Furthermore, the increased variability in allele frequencies after Omicron took over (after 1Q22) matches the broad peak in sequence diversity shown in 
Fig.~\ref{fig:num-of-seqs-per-week} (lower panel).

\subsection*{Fitness predicted by tQLE}
Epistatic fitness or covariation selection 
coefficients 
$f_{ij}$s
in QLE follow from the theory developed in
\cite{Kimura1965,NeherShraiman2011}
(in the version for bi-allelic loci).
This theory generalizes directly to 
multi-allelic loci~\cite{Gao2019,Zeng2020,Zeng2022},
and leads to the inference formulae eq.~(\ref{eq:f-2-inference}).
The additive fitness or selection coefficients $f_i$s 
are in tQLE
analogously
obtained   
as the differences between
the time derivative of an additive term ($\dot{h}_i$)
and a combination of the epistatic terms and
allele frequencies
($\sum_j f_{ij} m_j$). In the generalization to
multi-allelic loci, and when
time derivatives are approximated as 
discrete time differences, this 
leads to inference formula eq.~(\ref{eq:f-1-inference-QLE})
where $t$ and $t+\Delta$ stand for two different weeks.

\begin{figure}[!ht]
\centering
\includegraphics[width=0.47\textwidth]{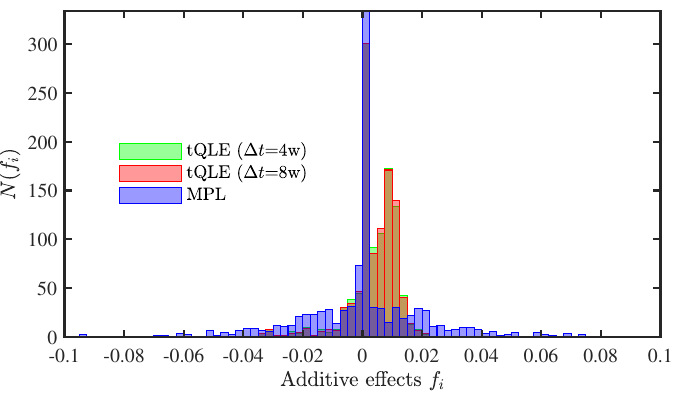}
\caption{ Histograms of the additive fitness from the tQLE (green and red bars) and MPL (blue bars) approaches. For the tQLE method, two distinct time intervals, namely $\Delta t = 4$ weeks (denoted by light green) and $\Delta t = 8$ weeks (denoted by light red), are considered in the additive term of equation \eqref{eq:f-1-inference-QLE}. This illustrative example showcases the outcomes for the week spanning from 2021-12-19 to 2021-12-25 of the UK dataset.}
\label{fig:histogram-tQLE-MPL}
\end{figure}

\begin{figure}[bt!]
\centering
\includegraphics[width=0.47\textwidth]{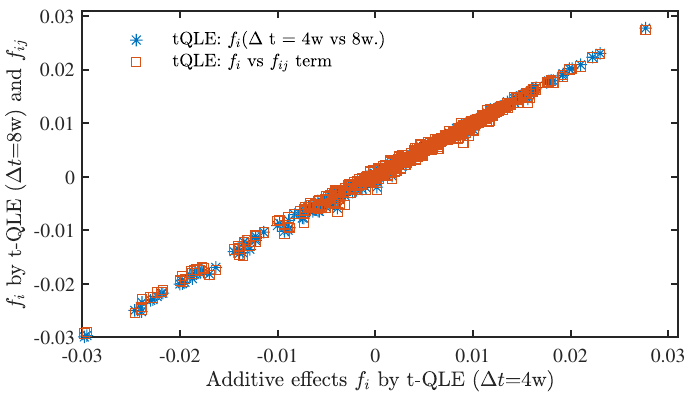}
\caption{Scatter plots for the additive fitness $f_i$ obtained from the tQLE. The presented results are derived from the dataset as of December 15, 2021, for the UK. Blue stars represent the $f_i$ values corresponding to a time interval of $\Delta t = 8$ weeks (on the y-axis) against those for $\Delta t = 4$ weeks (on the x-axis), while red squares represent the epistatic term of equation~\eqref{eq:f-1-inference-QLE} against the $f_i$ values for $\Delta t = 4$ weeks. Notably, both cases exhibit a close alignment with the diagonal, indicating a strong correlation between the compared terms.}
\label{fig:scatter-tQLE}
\end{figure}

As shown in Fig.~\ref{fig:histogram-tQLE-MPL}, the additive fitness by tQLE with the time interval of $\Delta t=4$ (green) and 8 (red) weeks are consistent with each other. 
This indicates that eq.~(\ref{eq:f-1-inference-QLE}) tQLE fitness values do not strongly depend on the choice of $\Delta t$ values, which is further shown in Fig.~\ref{fig:scatter-tQLE}.

Fig.~\ref{fig:scatter-tQLE} shows the epistatic terms versus $f_i$s inferred by the tQLE for $\Delta t=4$ weeks. The red squares lay closely around the diagonal. Such high consistency demonstrates that the epistatic term dominates over the time derivative $\dot{h}_i$ in the total inference formula for
$f_i$s. This pattern is consistent across all times in the present data set.   

\subsection*{Additive fitness inferred by MPL}
The additive fitness inferred by MPL validates its stability and reliability for the SARS-CoV-2 datasets. Here, even with a different stratification strategy for the UK datasets, similar results are obtained with those in~\cite{Lee2021}. 
While the great majority of fitness effects of mutations are inferred to be nearly neutral, MPL infers both strongly beneficial and deleterious mutations, as illustrated by the conspicuous deviation of the blue bars in Fig.~\ref{fig:histogram-tQLE-MPL} from the neutral zero point. 

\subsection*{Comparison between MPL and tQLE}
Fig.~\ref{fig:histogram-tQLE-MPL} shows the histograms of the additive fitness $f_i$s by the tQLE (green and red bars) and MPL (blue bars) models, respectively. The $f_i$s anticipated by the tQLE model exhibit a distribution proximate to the zero point, constraining in a relatively narrow range. In contrast, fitness effects inferred by MPL are mostly concentrated around zero, but with large deviations for a small number of mutations.

To assess the fitness estimates derived from the tQLE and MPL methods, we used both methods to rank sequences within a time window according to their fitness, and then compared these rankings. The fitness score of a sequence is defined in eq.~(\ref{eq:def-fitness-function}), in which the $f_i$s from the first term are the fitness effect while the $f_{ij}$s from the second term are the epistatic term.
The total fitness of a sequence from the MPL corresponds to the first term of eq.~(\ref{eq:def-fitness-function}) while that for the tQLE method is given by both terms of eq.~(\ref{eq:def-fitness-function}). For each
time window (one week)
the tQLE and MPL fitness scores for each 
sequence are computed and subsequently arranged in descending order. 

In Fig.~\ref{fig:SpearmanCorrelation} we show the Spearman rank correlation coefficient $c$ between the two rankings, stratified as to time (mean value per week, upper panel) and as to overall distribution (histogram, lower panel). 
The agreement is generally good 
($c>0.8$ for 105 out of 166 weeks). MPL and tQLE hence order sequences as to fitness in closely similar ways, for these UK-sampled SARS-CoV-2 sequences obtained from GISAID.

\subsection*{Change of agreement over time}
\begin{figure}[bt!]
\centering
\includegraphics[width=0.48\textwidth]{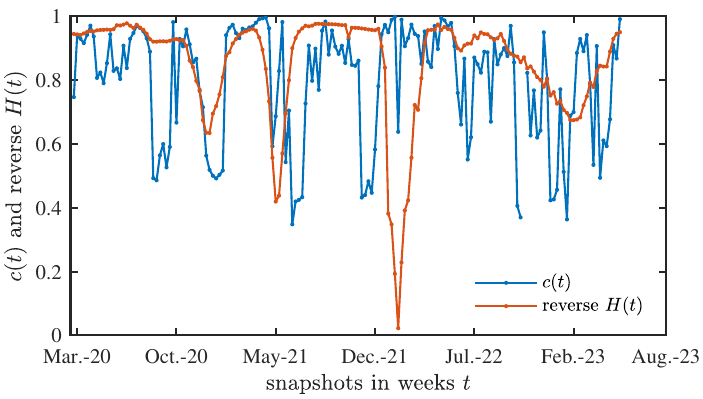}
\includegraphics[width=0.48\textwidth]{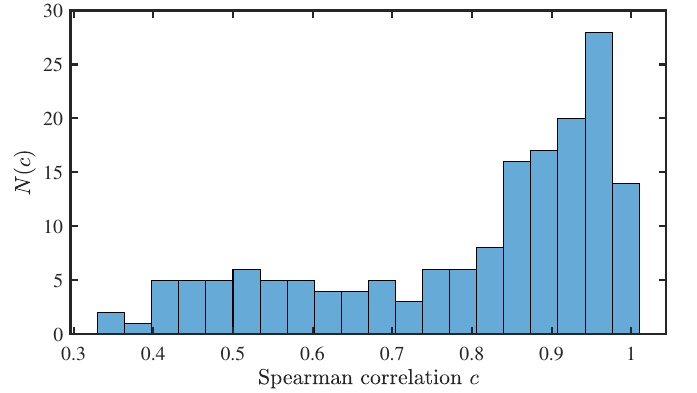}
\caption{
Upper panel: Weekly Spearman correlations $c(t)$ and relative Hamming distance deficit $rH^{(-)}$. (Blue curve) $c(t)$ for same UK data as in Fig~\ref{fig:num-of-seqs-per-week} (England, Scotland, Wales) from late February 2020 to early June 2023. First data point is for the time interval 2020-02-24 and 2020-03-14. 
(Red curve) $rH^{(-)}(t)=\left(H_{max}-H(t)\right)/H_{max}$ where $H(t)$ is the average Hamming distance between pairs of SARS-CoV-2 genomes sampled in week $t$ and $H_{max}$ (about 60 in this data set) is the maximum of $H(t)$ over all weeks. 
Lower panel: Distribution of the Spearman correlations $c$ between the top 10\% fitness provided by the MPL and the tQLE approach per week. In 105 out of 166 weeks $c>0.8$.}
\label{fig:SpearmanCorrelation}
\end{figure}

Fig.~\ref{fig:SpearmanCorrelation} also relates
agreement/disagreement in rankings to diversity (or lack thereof) in the sequences obtained in one week. 
For easier visualization we have plotted 
Spearman correlation $c(t)$ together with
the relative Hamming distance deficit 
$rH^{(-)}$ (defined in caption to Fig.~\ref{fig:SpearmanCorrelation}).
The two curves move in concert. This indicates a contravariation of
$c(t)$ with average Hamming distance $H(t)$, \textit{i.e.} that  
periods of high (low) Spearman correlation are related to periods of
low (high) sequence diversity.
A possible explanation
is that in this data set periods of high sequence variability (large average Hamming distance) appeared when one VoC was in the process of taking over the
population, see Fig.~\ref{fig:num-of-seqs-per-week} (lower panel).
At these times the UK viral population 
resembled a mixture of two clones,  
different from
the one Gibbs-Boltzmann distribution posited in QLE theory from which
the tQLE inference method has been derived, see Materials and methods.

\section{Discussion}
\label{sec:discussion}
Fitness is the central notion of population genetics going back to the beginning of the field. Inferring fitness has been cumbersome, and much of the literature has been dominated by theoretical investigations. The ongoing sequencing revolution has the potential to change this state of affairs, if fitness can be reliably inferred from large-scale population-wide whole-genome sequence data. 

We have here compared two approaches, MPL and tQLE, to infer fitness from time-stratified snapshots of an evolving population,
applied to UK data on SARS-CoV-2 during the COVID-19 pandemic\footnote{As discussed earlier, UK data is particularly suitable to this comparison due to the homogeneity of care and sequencing across this limited geographic area.}.
Comparisons of the inferred fitness of SARS-CoV-2 sequences during discrete time windows reveal a varying level of correlation between the two approaches.
During a large fraction of time windows the agreement between the two approaches is quite strong, as shown in Fig.~\ref{fig:SpearmanCorrelation}.

The gold standard for the accuracy of inferred fitness is comparison to experiments.
Nevertheless, different inference schemes can be compared between themselves, and inter-scheme agreement is a proxy when experiments are not available, as they are not on a
population-wide and global genomic scale for SARS-CoV-2.
The two approaches rest on simplifying assumptions of different kinds. In MPL, the fitness landscape is assumed to contain only additive components of fitness, which is known to be a simplification. In tQLE, on the other hand, the instantaneous state of the population is assumed to be in a Quasi-Linkage Equilibrium state, \text{i.e.}, as in a Gibbs-Boltzmann distribution with effective energy terms dependent on fitness. It is not \textit{a priori} clear which of the two sets of assumptions is the strongest for a strongly recombining virus like SARS-CoV-2, and their relative strengths could also have varied during the pandemic. 

We posit that for future pandemics sequence data is very likely to be abundantly available and available much sooner than experimentally determined fitness scores. Fitness parameters systematically inferred from data may then yield predictions useful in the analysis and the understanding of how the pandemic evolves, and to the choice and evolution of counter-measures.
Furthermore, our results suggest the possibility of combining the two methods by taking the stochastic dynamics in a QLE state as developed in \cite{NeherShraiman2011} into account. Recently, MPL was extended to consider epistatic interactions \cite{sohail2022inferring}, but the resulting expressions are computationally intensive and require the calculation of fourth-order correlations. This has made epistatic inference with MPL challenging for populations with large numbers of mutations, as observed for SARS-CoV-2. Introducing ideas from QLE could then reduce the computational burden and widen the scope of MPL.


\bibliography{Covid19}

\end{document}